# Revisiting literature observations on photodarkening in $Yb^{3+}$ doped fiber considering the possible presence of Tm impurities

Romain Peretti[1], Cédric Gonnet[2], and Anne-Marie Jurdyc[1]*

*1-Université de Lyon, Université Lyon 1, CNRS/LPCML, Villeurbanne F-69622, France*
*2-Draka, Data center IV, Route de Nozay, 91460 Marcoussis, France*

## ABSTRACT

Ytterbium (Yb) doped fiber lasers are known to be affected by the creation of color centers during lasing (so-called photodarkening (PD)). In a previous work, this defect creation was investigated from a spectroscopic point of view, showing the presence of traces (ppb) of thulium (Tm) in the Yb doped fiber. It was shown that Tm has a strong impact on the defect creation process involved in PD. In this paper, we compare the results from the literature with our Tm hypothesis, without finding any contradiction. Moreover, this hypothesis can be an explanation for the discrepancies in the literature.

## INTRODUCTION

Following the first laser emission [1], and the steps leading to the development of modern optical fibers [2, 3], these two pillars of photonics were finally married by Stone and Burrus in 1973 [4, 5] when they built a fiber laser, even if the phrase had already been used to describe what is now known as a rod-type fiber laser [6]. Combining the good modal quality of a single mode fiber with a long amplification medium leads to a powerful optical pumped laser with extremely good modal qualities, which can be a reliable solution when building high power density lasers [7] (more details on fiber lasers can be found in [8, 9]). There are many applications for such lasers, including military uses, medical uses, marking, etching, engraving and trimming [10, 11, 12]. Many doping elements can be used to make a fiber laser, especially the rare earth elements ($Pr^{3+}$, $Nd^{3+}$, $Tm^{3+}$, $Ho^{3+}$, $Nd^{3+}/Yb^{3+}$, $Nd^{3+}/Yb^{3+}$)[8], including the well-known erbium that is used for telecommunication amplifiers (the Erbium Doped Fiber Amplifier (EDFA) [13]). But nowadays the most promising fiber laser is the Yb doped fiber laser [14]. Since its first demonstration in the late 1980s by Suni, *et al*. [15, 16], the commercially available output power for this laser has reached 10 kW for a continuous wave, and 50 kW in pulsed mode. Such power allows applications in areas such as machining or surgery, but could also be used in non-linear optics in order to provide an extreme ultraviolet source for lithography for the next generation of microprocessors [17].

However, the qualities required from a laser are not only power and reliability but also durability. In fact, in 1997 Pashotta, *et al*[18] discovered that the laser's output power and threshold deteriorated during the running of a $Yb^{3+}$ doped fiber laser. This degradation is attributed to photodarkening (PD). The effect of this is that a broad absorption band, from ultraviolet to near infrared, appears during laser operation. So-called PD had already been reported for other rare earth doped fibers ($Pr^{3+}$ [19], $Eu^{2+}$ [19], $Tm^{3+}$ [20, 21], $Ce^{3+}$ [22], $Tb^{3+}$ [23], $Ta^{3+}$ [24]) and was reported in $Sm^{3+}$ [25] later. PD is also known to occur in other applications such as non-linear optics [26, 27, 28], and in other propagative materials like polymer fibers [29]. For Yb, PD was also reported for matrices other than optical glass fiber [30].

The unusual feature of PD in Yb doped fiber lasers is the creation of an absorption band at higher energy than that of the laser's pump or signal, and higher energy also than that of the levels of active ion [31]. In fact, it is well known that high energy radiation induces the creation of an absorption band in a matrix like silica glass [32, 33], and UV light is also used to make Bragg mirrors into optic fibers [34], even with a UV lamp [35], but IR light is not supposed to do the same.

The absorption band created by PD is attributed to the creation of color centers; this could be due to charge transfer [36], non-bridging Oxygen Holes [37], oxygen deficiency centers (ODC) [38], $Yb^{2+}$ ions [36, 39, 40], or even $Yb^{2+}$-$Yb^{3+}$ pairs [41]. Analysis of defects created by high energy irradiation in silica glass is a large and complex subject which is outside the scope of this review (for further detail, see [32]). In this review, we will focus on the observations made in the literature about the process leading to the creation of the defects, the so-called PD, and not on the defects themselves.

To analyze the PD in $Yb^{3+}$ doped fiber lasers, Koponen, et al.[42] designed an experiment in order to measure the absorption created during laser irradiation. It consists of measuring the absorption at a fixed wavelength (633 nm, for example) after a given time of exposure to a PD beam (usually of 920 nm or 976 nm). To avoid measurable darkening beam depletion, the researchers chose to reduce the length of the analyzed fiber to a minimum, corresponding to around 10 cm. The result of this experiment is often depicted as a plot of absorption (or transmission) against time of exposure. In addition, this curve is often fitted to a stretched exponential decay curve [43, 44], such as is often used to depict solid-state relaxation phenomena associated with a microscopic reversible process, as described in (1).

$$\alpha(t) = \alpha_{st}\left[1 - \exp\left(-\left(\frac{t}{\tau}\right)^{\beta}\right)\right] \quad (1)$$

where $t$ is the time's abscissa, $\alpha_{st}$ is the absorption at the steady state (for $t \gg \tau$), $\beta$ is the stretching parameter and $\tau^{-1}$ is the PD rate.

Although the mathematical relationship between the parameters $\alpha_{st}$, $\beta$ and $\tau^{-1}$ of (1) has not yet, to our knowledge, been clearly reported in the literature, $\tau^{-1}$ nicely represents the sensitivity of a fiber to PD. The higher the $\tau^{-1}$ parameter, the lower the sensitivity of the fiber to PD, and the greater its lifetime. Therefore this parameter is often used to compare the resistance of fibers to PD. The stretched exponential fit is a useful tool for global studies, but we should be careful about using it in more precise analyses.

Many papers and conferences have reported results with similar set-ups and analyses in the study of PD dependence on different parameters, such as laser power or population inversion, or different characteristics of the Yb doped materials such as the Yb concentration, the matrix composition, or the role of co-doping ions.

Recently, in our work [45, 46, 47] we analyzed in depth the UV and visible emission from an ordinary Yb doped fiber, and showed emissions corresponding to four transitions of $Tm^{3+}$ ions ($^1G_4 \rightarrow ^3F_4$ at 650 nm, $^1G_4 \rightarrow ^3H_6$ at 475 nm, $^1D_2 \rightarrow ^3H_6$ at 370 nm and ($^1I_6$, $^3P_0$) $\rightarrow ^3H_6$ at 310 nm) [48, 49, 50, 51] under 976 nm excitation. $Tm^{3+}$ emission had already been observed at 475 nm by Kirchhof, et al[52]. It is known that $Tm^{3+}$ ions can easily up-convert infrared light to visible or UV light [48, 53], especially when they are coupled with $Yb^{3+}$ ions [54]. Therefore we decided to add 300 ppm of Tm to our $Yb^{3+}$ doped fiber, and we showed that it strongly increased the sensitivity of our sample to PD. In summary, the two UV emissions of $Tm^{3+}$ could be absorbed by the matrix, and in addition Tm is known to give rise to strong PD [20, 21]. So we conclude that contamination of the raw Yb by Tm ions should be the indicator for PD. The first idea arising from this conclusion was to remove the Tm completely from the raw Yb. Unfortunately, to our knowledge, it is not possible to have Yb material purer than 99.9999% (6N), so it is not possible to remove Tm completely from Yb doped fibers. The most important corollary of this conclusion is that, to avoid PD, two steps must be taken: the first being to prevent $Tm^{3+}$ ions from reaching high energy levels, and the second to prevent these high energy levels impacting on the absorption of the fiber.

Soon after this, Jetschke, et al. [55] conducted precise analyses of the Tm co-doping dependence on the PD kinetics in Yb fibers. They concluded that, if Tm-induced PD exists, then intrinsic Yb doped fiber PD also exists and can predominate when the contamination is very low.

In the following, we review the observations reported in the literature regarding the contamination of Yb raw material by Tm ions, and suggest that this should be the indicator for PD. Because all the thermal studies of PD are focused on the defects created during the process and not on the process itself, and so $Tm^{3+}$ induced PD will not change the interpretation of any of these studies, we decided not to focus on these observations.

# DEPENDENCE ON THE SOURCE WAVELENGTH

In the literature, PD is shown to occur when $Yb^{3+}$ doped fibers or preforms are pumped at different wavelengths or powers. We will first look at the effect of the darkening wavelength and then at the darkening power.

After reporting PD at the pump (920 nm or 976 nm) and signal (from 1µm to 1.2 µm) wavelengths, other analyses at different darkening wavelengths were made in order to obtain a better understanding of the PD process. PD was also shown to occur at 488 nm [38, 56], with UV laser excitation [57], and even with a UV lamp [36, 58]. Notice that in all cases the reported absorption spectra are similar to those burned around 1µm [38, 59]]. As can be seen in **FIGURE 1**, PD at 488 nm is consistent with there being $Tm^{3+}$ contamination. In fact, the $^1G_4$ level of $Tm^{3+}$ is known to exhibit a strong absorption band at around 488 nm, and is able to up-convert 488 nm radiation to its UV level $^1I_6$-$^3P_0$ by one Excited State Absorption (ESA) [60]. This population at high energy levels is able to emit UV light and subsequently create defects in the same way as UV lasers and lamps [35] create defects in silica glass, especially in the presence of $Tm^{3+}$, which is used as a sensitizer [61, 62].

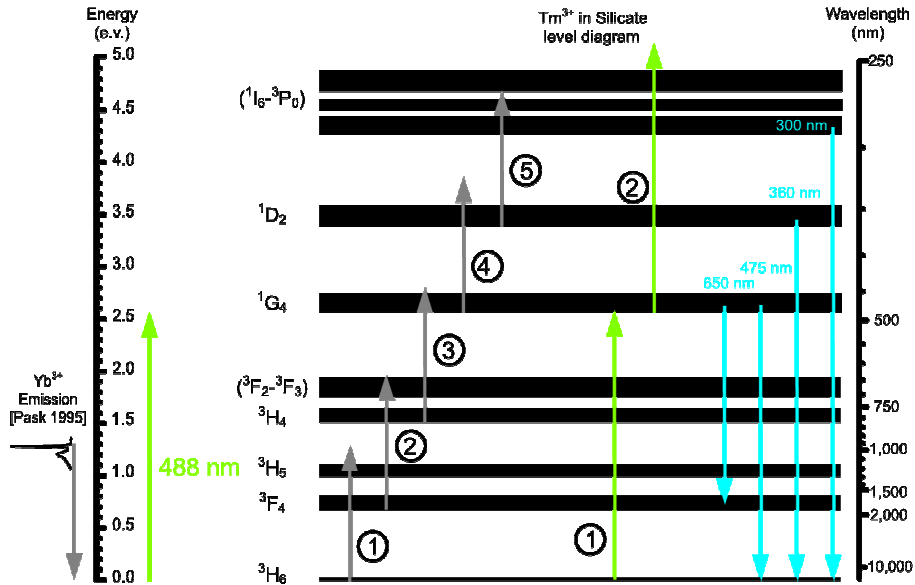

**FIGURE 1** : Schematic of the $Tm^{3+}$ and $Yb^{3+}$ energy levels and proposal for the process leading to UV emission in $Yb^{3+}$ doped fiber contaminated by $Tm^{3+}$ (grey arrows represent the signal around 1 µm, and green the burning signals around 488 nm, IR Yb emission from Pask [35])

# DEPENDENCE ON THE SOURCE POWER

In addition to the wavelength, the light source's main parameter is its power density. In fact, large mode area (LMA) fibers [63], developed to decrease nonlinear effects by reducing power density, were good laboratory objects for analyzing PD in the case of relatively low power density, and good components for reducing PD [64]. However, with the increasing demand for power, this solution reached its limits [65]. Furthermore, even though PD was known to be dependent on the power [66], studies on this dependence were only recently carried out [67, 65, 68, 69].

Next, we will first describe some theoretical clues for understanding the results from the literature, and then we will give an explanation of these results. First of all, most authors prefer to refer to an inversion parameter instead of pump power or pump power per unit surface. This inversion parameter is the ratio of $Yb^{3+}$ ions in the excited state and, in the absence of a laser signal, can be expressed by (2):

$$N_{^2F_{5/2}} = \frac{\sigma_{abs} \times I_{pump} \times \tau_0}{h\nu_{pump} + \sigma_{abs} \times I_{pump} \times \tau_0} \Leftrightarrow I_{pump} = \left(\frac{h\nu_{pump}}{\sigma_{abs} \times \tau_0}\right) \times \frac{N_{^2F_{5/2}}}{1 - N_{^2F_{5/2}}} \tag{2}$$

where $I_{pump}$ is the mean power density of the pump (W.m$^{-2}$), $\sigma_{abs}$ is the absorption cross-section of the $^2F_{5/2}$ level of Yb$^{3+}$ at the pump frequency $\nu_{pump}$ and $\tau_0$ is level's lifespan. To obtain the inversion formula during lasing, $\tau_0$ should be replaced with $\left(\frac{1}{\tau_0} + \frac{\sigma_{em} \times P_{signal}}{h\nu_{signal} \times A_{signal}}\right)^{-1}$. With that equation we can understand the simple homographic relationship between pump power ($I_{pump}$) and inversion rate ($N_{2F_{5/2}}$).

We note that phenomena involving excited rare earth ions, such as energy transfer (ET) or cooperative emission (CE), should have a probability proportional to $[N_{excite}]^n$, with $n$ being the number of excited ions involved, and that phenomena involving only one ion but more than one photon, such as absorption, excited state absorption (ESA) and two-photon absorption (TPA) [70], should have probability proportional to $[I_{pump}]^m$, where $m$ is the number of photons involved.

Most papers describe the PD power dependency using a graphic depicting the PD rate $\tau^{-1}$ of (1) versus the inversion rate, and not versus the pump power, and neglect ESA and TPA, which cannot occur in Yb$^{3+}$ doped fibers, and also additional defects or impurities. As they do not consider processes involving several photons, but only those involving several ions, the data are fitted by the straight line equation given in (3).

$$\tau^{-1} \propto [N_{2F_{5/2}}]^n \qquad (3)$$

The $n$ parameter in (3) describes the number of excited Yb$^{3+}$ ions involved in the PD mechanism.

However, if we consider a fiber contaminated with another rare earth such as Tm$^{3+}$, not only ET and CE, but also ESA and TPA, should be taken into account. In fact, we have shown in previous publications [46, 47] that Tm emits in Yb doped fiber at different wavelengths. One can note that the $^1G_4 \rightarrow ^3H_6$ emission of 210 ppbw of Tm$^{3+}$ (475 nm) is about the same intensity as the cooperative emission of Yb$^{3+}$ associated with 1.7 wt.% of Yb. This effect has also been observed by Goldner, et al.[71]. The cooperative emission efficiency is very low, around 10$^{-8}$ cm$^2$/W [72], even if this emission is plainly visible to the naked eye. The mechanisms leading to these emissions have been analyzed, and it appears that at low Tm$^{3+}$ concentration a five-step Yb-Tm ET takes place [73], leading to the successive population of the $^3H_5$, $^3F_2$, $^1G_4$, $^1D_2$ and finally the $^3P_0$-$^1I_6$ energy levels. In addition, it was shown in [73] that increasing the ratio of the concentration of Yb$^{3+}$ to that of Tm$^{3+}$ leads to an increase in the populations of the $^1I_6$ and $^1D_2$ levels, and therefore an enhanced emission in the UV. The ESA up-conversion phenomenon is so efficient in Tm$^{3+}$ that a laser at 284 nm was demonstrated [74]], showing that one cannot ignore these processes.

In such cases the PD rate $\tau^{-1}$ would be strongly dependent on both light power density and inversion rate. The $\tau^{-1}$ parameter in PD can be expressed by (4), using the approximation that up-conversion processes are less probable than emission ones.

$$\tau^{-1} \propto [N_{2F_{5/2}}]^n \times [I_{pump}]^m \propto \frac{[N_{2F_{5/2}}]^{n+m}}{[1 - N_{2F_{5/2}}]^m} \qquad (4)$$

In (4), the $n$ parameter describes the number of excited Yb$^{3+}$ ions involved in the process, and the parameter $m$ depicts the number of photons involved in the process. We can see that at low inversion rates, when $N_{2F_{5/2}} \ll 1$, (4) is similar to (3), with $n$ replaced by $n+m$, but when the inversion rate is high, so $N_{2F_{5/2}}$ is close to 1, the expression in (4) is far greater than $[N_{2F_{5/2}}]^{n+m}$. That means that if such a curve is fitted by an exponential model as we did for (3), the value of $n$ that is found will be strictly greater than $n+m$ and the higher the inversion, the greater the overestimation of $n$ above the $n+m$ value.

In the literature, the reported values of the $n$ parameter range from 2 [67] to 7 [69, 75]. Indeed, two groups can be defined:

Below 5: in [67] ($n$ =2 for λ=976 nm and inversion from 0.1 to 0.5), [68] ($n$ =3.5±0.5 for λ=915 nm and inversion from 0.1 to 0.6, independent of the Yb$^{3+}$ concentration), and [44]: ($n$ =4.3±0.5 for λ=915 nm and inversion from 0.35 to 0.6); and Above 5: in [76] ($n=7\pm0.5$ for $\lambda=920$ nm and inversion from 0.25 to 06), [69] (with two values (long and short time) for the two fibers studied $n_{long}=7.7\pm1$, $n_{short}=7.2=\pm0.8$ and $n_{long}=6.4\pm0.6$, $n_{short}=6.9\pm0.6$ for $\lambda=976$ nm and inversion from 0.2 to 06).

Notice that some studies present their results according to the samples concentrations – for example [68] shows a quasi non-dependence of the $n$ parameter on concentration. These results are plotted in **FIGURE 2** in order to show these 2 groups.

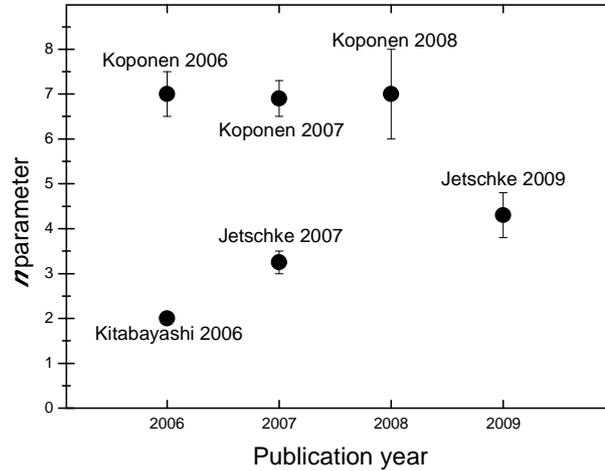

**FIGURE 2** : $n$ parameter in the literature (Koponen 2006 [42], Kitabayashi 2006 [67], Koponen 2007 [65], Jetschke 2007 [68], Koponen 2008 [69], Jetschke 2007 [44])

All the results from the literature came from model which ignores ESA and TPA. However, if one fitted phenomenon involving ESA or TPA onto a straight line in a log-log scale, the value found for $n$ would be overestimated. This could be an explanation for the discrepancy in the published results regarding the value of $n$, since the published $n$ parameters would be overestimated first because the value ($n+m$) was taken, and then because, at high inversion rates, the slope in (4) gives an overestimated value for the ($n+m$) parameter. This reasoning can especially be applied to results above $n=6$. From this, we conclude that ESA and TPA cannot be ignored in PD processes. This, as well as the influence of the matrices, could explain the discrepancy between the results in the literature. Even if the discrepancy shows that the phenomenon is not yet well-known, the latter papers clearly exhibit a multi-step process similar to the one we proposed in our work involving $Tm^{3+}$ [45, 46]. Jetschke and Röpke [44] concluded with a process involving four to five excited $Yb^{3+}$ ions. Their result agrees with our conclusions in [46]. Even though the process of up-converting energy from the 976 nm pump to the UV levels of $Tm^{3+}$ is not yet fully established, we give an example in **FIGURE 1** for a possible five-step path for reaching the highest UV energy band of $Tm^{3+}$ and a possible four-step path for reaching the lowest energy band. Each step can be ET from a $Yb^{3+}$ donor to a $Tm^{3+}$ acceptor or ESA of $Tm^{3+}$. We cannot determine which of these is the more probable in our samples; however the probability should strongly depend on the $Yb^{3+}$ concentration. The greater the $Yb^{3+}$ concentration, the more probable is ET.

In addition, photobleaching, which consists of the irradiation of a pre-darkened sample with light of a given wavelength in order to reduce or remove PD, is observed (using irradiation at 980 nm in [68, 41], at 540 nm in [41] and at around 355 nm in [77]). Photobleaching could also be explained by the contamination with $Tm^{3+}$ ions: indeed bleaching with UV photons has been known for a long time to take place [78] in silica glasses, and we have shown that UV emission exists in $Tm^{3+}$ contaminated $Yb^{3+}$ doped fibers when they are irradiated at 980 nm or 540 nm.

To summarize, we are proposing an up-conversion scheme that entails a mix of a four- and a five-step process for PD. Each step of this up-conversion could be ESA or ET, and should depend on the sample properties (concentration, clustering, spectral overlap…). In our work [45, 46], if PD is related to the 310 nm emission band then $n+m=5$ in (4), and if it is related to the 370 nm band then $n+m=4$. Notice that $n+m$ corresponds to the total number of steps, with $m$ corresponding to the number of ESA steps (which should depend on the number of photons), and $n$ corresponding to the number of ET steps (which should depend on the number of excited $Yb^{3+}$ ions).

For future work, it should be interesting to obtain a better understanding of this power and inversion dependency on PD. One way of proceeding would be to look at the dynamics of all these processes in order to distinguish them. In addition, an indication should be given by analyzing the power and $Yb^{3+}$ inversion dependency of the UV emission of $Tm^{3+}$ by using the same kind of pulsed laser. Moreover, studying the whole dynamic process of up-conversion leading to $Tm^{3+}$ UV emission, by measuring the lifespan, could also be a clue to a better understanding of PD and to avoiding PD in $Yb^{3+}$ doped fibers.

## MATERIAL EFFECTS

Optimizing the optical glass material of the $Yb^{3+}$ doped fiber is one of the possibilities for reducing or even avoiding PD. That is why many papers report on the influences of the $Yb^{3+}$ concentration, the co-doping impact and the core composition effects on PD.

### Ytterbium concentration

Because $Yb^{3+}$ concentration is a key parameter for laser or amplification applications, the effect of $Yb^{3+}$ concentration on PD has been studied [67, 68, 57, 69]. Two methods were applied when conducting these studies.

The first method consists in recording PD losses for fibers with different concentrations but with a fixed inversion rate. In that way, Kitabayashi, et al [67] found a PD losses dependence law with concentration in: concentration raised to the power of 2.69 or 2.64(varying with Al concentration); and PD losses dependence law with burning power in: burning power raised ton the power of 2. Those imply that more than one ion is needed for the processes, but one cannot, from this, identify the processes themselves or even the number of processes involved. Other studies found that a more accurate parameter to be taken into account is the concentration of $Yb^{3+}$ ions in the excited state, the so-called [Yb*].

The second method consists in recording PD losses for the same concentration but with different inversions. Using this method, similar results were found [68, 69] for the dependence of PD losses on inversion to those for the dependence on excited $Yb^{3+}$ concentration. To summarize, the second method, identifying the influence of concentration and power together, appears from the literature to be a good protocol for comparing fibers with the same composition but different concentration, and all the studies conclude that more than one ion is needed for the process. Moreover, in [68], for four out of five studied fibers, at a given [Yb*] the higher the concentration of $Yb^{3+}$ in the fiber, the more rapid is the PD and the more important are the losses at equilibrium.

The conclusion from these studies, that more than one $Yb^{3+}$ ion is involved in PD, is in good agreement with the mechanism we proposed in [45, 46]. Actually, increasing the $Yb^{3+}$ concentration would also increase the $Tm^{3+}$ contamination, which could be an explanation for the growth in PD with $Yb^{3+}$ concentration, and thus $Tm^{3+}$ contamination, in [68]. To better understand the influence of concentrations, Jetschke, *et al.* [55] show an increase in PD versus $Tm^{3+}$ concentrations at a fixed $Yb^{3+}$ concentration, for $Tm^{3+}$ concentrations greater than 65 ppbw. In the future it would be interesting to analyze PD dependence on $Yb^{3+}$ concentrations at a fixed $Tm^{3+}$ concentration.

### Co-doping effect

One way to reduce or avoid PD could be to add a competing phenomenon. With that purpose in view, two strategies could be applied.

The first would be to reduce defect creation resulting from the interaction of fiber material with high energy photons. Engholm et al [79, 80, 81] co-doped $Yb^{3+}$ doped fiber with cerium. Because of its strong oxidizing properties, $Ce^{4+}$ [82] is commonly used in the glass industry as a decolorizing agent and so could reduce defect creation. The researchers reported a strong reduction of PD loss in Ce co-doped $Yb^{3+}$ fibers. They explain this by saying that, because of the presence of both $Ce^{3+}$ and $Ce^{4+}$ in the fiber, cerium could serve as a trap for free electrons, or for free holes, that can form color centers in the PD process. In that case, because the energy levels of both $Ce^{3+}$ and $Ce^{4+}$ are far from those of $Yb^{3+}$, energy transfer would not have great consequences on the laser itself.

The second strategy would be to avoid high energy levels being reached by adding a de-excitation channel by, for example, co-doping with $Er^{3+}$. Morasse et al[59] demonstrated that co-doping a $Yb^{3+}$ doped fiber with $Er^{3+}$ greatly reduced PD, and this result was shown again by Jetschke, *et al.* [55]. Gavrilovic, et al. [83] gave hints about the ability of $Er^{3+}$ to reduce PD in $Tm^{3+}$ contaminated $Yb^{3+}$ fibers. In fact, the well-known de-excitation channels from the $Tm^{3+}$ levels $^3F_4$, $^3H_5$ and $^3H_4$ to, respectively, the $Er^{3+}$ levels $^4I_{15/3}$, $^4I_{13/2}$ and $^2H_{3/2}$ or $^2H_{5/2}$ will prevent $Tm^{3+}$ from reaching higher energy levels. However, $Er^{3+}$ co-doping will strongly impact on the laser characteristics of the fiber. Other co-doping elements were proposed in [83], such as $Tb^{3+}$, $Eu^{3+}$ and $Nd^{3+}$, but in each case the laser characteristics of the fiber need to be analyzed carefully in order to ensure that the other qualities of the $Yb^{3+}$ doped fiber laser are not lost.

In conclusion, both the strategies, that of co-doping in order to avoid $Tm^{3+}$ reaching high energy levels and that of co-doping in order to suppress defect creation, could be applied. The former should carefully take laser properties into account. Both strategies should be a good compromise for greatly reducing PD losses in $Yb^{3+}$ doped fibers without damaging the laser properties.

## Matrix effect (core composition)

The final key point in the role of the material on PD is the glass composition itself. The influence on PD of the usual co-dopant for silica based optical fibers, like aluminum, germanium or phosphorus, has been extensively studied. In 2006 Kitabayashi et al[67] demonstrated that aluminum is effective at reducing PD losses. Actually, adding aluminum allows the $Yb^{3+}$ concentration to be increased by a factor of 4 while maintaining the same level of PD losses. That conclusion on the effect of aluminum co-doping was confirmed in 2007 by Morasse, et al[59], but the addition of germanium did not exhibit any effect on PD. In 2008, Engholm and Norin [58] confirmed the role of aluminum, but reported that phosphorus co-doping was even more efficient at reducing PD, and insisted that there was a correlation between UV absorption of the host matrix and PD sensitivity. Lee, et al[57] showed that adding phosphorus to a fiber allowed an increase in $Yb^{3+}$ concentration by a factor of 6 while keeping the same level of PD. In the same year Jetschke et al [84, 85], and, later, Suzuki, et al. [86], studied the addition of both phosphorus and aluminum, and proposed an optimized composition to reduce PD. In 2009, a mode-locked $Yb^{3+}$ doped phosphate fiber laser was reported to have low PD [84]. Other compositions, such as a Yb doped $Y_2O_3$ nanoparticles fiber, were shown to exhibit very low PD [87].

First, the correlation between the UV matrix absorption and PD [58] could well be explained by the presence of a real level in the vicinity of the matrix absorption band, increasing the probability of creating additional defects. This real level could be any defects or impurities level such as the $Tm^{3+}$ levels. This observation is consistent with the $Tm^{3+}$ contamination effect.

Secondly, the use of aluminum and phosphorus [88, 89] is well-known for avoiding the multi-ions process in rare earth doped silica glass, so the phenomenon has been well-studied, especially for erbium ions for telecommunication purposes. Monteil, et al[90]] have shown that most of the rare earth is located close to the aluminum. The vicinity of the aluminum affects the spectral properties of the ions, avoiding interaction. This effect was confirmed in [91, 92] and was microscopically studied by Saitoh, et al. [93]. In this paper the authors also showed that the effect of phosphorus was to induce a cage around a part of the rare earths, limiting, clustering and modifying their spectrum, as shown in [94]. Focusing on the $Yb^{3+}$ ions, Deschamps, et al. [95] confirmed that phosphorus is more efficient at avoiding clustering than aluminum, and Oppo et al. [96] showed by X-ray absorption spectroscopy that $YbPO_4$ nanocrystals are present in Yb doped fiber.

The mechanism proposed here for $Tm^{3+}$ induced PD will be strongly dependent on clustering. All the literature observations are consistent with this mechanism. Practically, clustering would enhance energy transfer from $Yb^{3+}$ to $Tm^{3+}$, and thus would increase PD. Another technique, proposed by Koponen et al. [66] and Tammela et al. [97], could be useful for mitigating clustering and consequently PD. It consists of doping the silica fiber not directly with rare earth, but with an amorphous nanoparticle, allowing a better control of the rare earth environment and dilution and therefore of clustering.

## CONCLUSION

To summarize, we have considered the literature relating to PD in Yb-doped fibers, looking at the darkening source, the thermal or bleaching effects and the dopants of the fiber. In all cases the observations were not inconsistent with the role of $Tm^{3+}$ in PD, and did not allow us to discriminate intrinsic PD from $Tm^{3+}$ induced PD. To be exact, our explanation is consistent with the coexistence of UV, 488 nm and 980 nm PD, and with the increase of PD with pump power or $Yb^{3+}$ concentration. In addition, most of the studies on material focus on the role of the matrix on the interaction of ions and energy transfer, saying that the less often these phenomena occur, the less PD there is. Moreover, this hypothesis is a clue to explaining the discrepancies in the literature concerning the dependence of PD on inversion.

We showed from the literature that up-conversion of $Tm^{3+}$-$Yb^{3+}$ can lead to UV emission, UV that is known to create PD in silica fiber. To be precise, we were able to measure [46] UV emissions in an alumino-silicate fiber doped with Yb precursor of 4.5N purity (0.99998% pure).

To conclude, we reviewed the literature looking for a contradiction to our hypothesis on the role of $Tm^{3+}$ impurities in PD, and we were not able to find any contradiction. This absence of contradiction implies that our hypothesis is reinforced. To go further in the study of PD, researchers must at least give the purity of the Yb precursor used to dope the fiber or, even better, must analyze their fibers to report the $Tm^{3+}$ concentration, as in [55].

If the purity of the material is not high enough, it may be possible to reduce the interaction of the ions by adding other elements, such as cerium, which are known to trap high energy, and thus to reduce the UV sensitivity of the matrix. The next step could be to place $Ce^{3+}$ ions close to the $Yb^{3+}$ or $Tm^{3+}$ ions in order to optimize the trapping effect. The main focus should be to avoid UV emission from Tm by quenching the up-conversion process, as proposed in [83], but to maintain the good quality of the $Yb^{3+}$ doped fiber laser. To reach that goal, doping with other rare earths, nanoparticles or even molecules like $OH^-$ could greatly help.

Finally, we will again insist on the importance of contamination of the laser raw material when really high optical intensity is reached because, even with ppb concentration, these contaminations could be activated by high electro-magnetic fields. In $Yb^{3+}$ doped fiber, a precursor purer than 5N (99.999% pure) must be used.

## ACKNOWLEDGEMENTS

The authors want to thank Ekaterina Burov, Alain Pastouret, Bernard Jacquier and Florian Kulzer for their fruitful discussions, and the reviewer for help with improving the manuscript.